# P-type polar transition of chemically doped multilayer $MoS_2$ transistor

*Xiaochi Liu [†], Deshun Qu [†], Jungjin Ryu, Faisal Ahmed, Zheng Yang, Daeyeong Lee, and Won Jong Yoo [*]*

X. Liu, D. Qu, J. Ryu, F. Ahmed, Z. Yang, D. Lee, Prof. W. J. Yoo
Samsung-SKKU Graphene/2D Center (SSGC)
Department of Nano Science and Technology
SKKU Advanced Institute of Nano-Technology (SAINT),
School of Mechanical Engineering
Sungkyunkwan University, 2066, Seobu-ro, Jangangu, Suwon, Gyeonggi-do, 440-746, Korea.
E-mail: yoowj@skku.edu

[†]These authors contributed equally to this work

**Abstract**

The accessibility of both n-type and p-type $MoS_2$ FET is necessary for complementary device applications involving $MoS_2$. However, $MoS_2$ PFET is rarely achieved due to pinning effect resulting high $R_c$ at metal-$MoS_2$ interface and the inherently strong n-type property of the $MoS_2$ material. In this study, we realized a high-performance multi-layer $MoS_2$ PFET via controllable chemical doping, which has an excellent on/off ratio of $10^7$ and a maximum hole mobility of 72 $cm^2$/Vs at room temperature, and these values are further exceeding to $10^9$ and 132 $cm^2$/Vs at 133K. In addition, we revealed that large $R_c$ hindered the polar transition of $MoS_2$ FET from n-type to p-type, meanwhile channel $R_s$ limited $I_{on}$ of PFET. Therefore it is suggested that reducing $R_c$ at high work function metal-$MoS_2$ interface and p-type doping of channel were necessary for achieving high performance $MoS_2$ PFET. Based on the high performance PFET, we successfully demonstrated a $MoS_2$ CMOS inverter by integrating NFET and PFET.

**Key words:** $MoS_2$; P-type; contact resistance; transistor; chemical doping

# 1. Introduction

Over the past few years, two-dimensional (2D) transition metal dichalcogenides (TMD) have attracted enormous attention due to their unique material properties and potential applications in various electronic, optical, and spintronic devices;[1-6] however, integration of n-type and p-type field effect transistors (NFETs and PFETs) needs to be developed to enable complementary device operations using TMD. Several methods, including metal work function engineering, chemical doping, electrostatic doping, and ionic gating, have been tested for their utility in achieving bipolar carrier conduction in TMD.[7-10] These experiments were mostly carried out using 2D tungsten diselenide ($WSe_2$), in which the Fermi level lies in the middle of the band gap and no pinning occurs at the metal–$WSe_2$ interface, enabling relatively easy tuning of polarity.[8] Molybdenum disulfide ($MoS_2$) is widely studied as a typical 2D semiconductor, however, elemental metals, including high work function metals, such as Pd, Ni, and Au, give rise to n-type $MoS_2$ transistor properties when used as source and drain contacts.[11-13] Most studies have ascribed the stubborn n-type behavior to Fermi level pinning at the metal–$MoS_2$ interface.[14-16] The employment of a high work function $MoO_x$ material as a hole injection layer at the contact interface was reported to successfully provide a $MoS_2$ PFET;[17] however, ultra-high vacuum conditions and high temperatures are needed for the deposition of $MoO_x$, and the on-current density ($I_{on}$) of the fabricated $MoS_2$ PFETs was low. As the drain bias or gate bias increased, the electron current appeared to increase, yielding slight ambipolar behavior. Other studies have tested the utility of Nb as a substitutive dopant introduced into both mechanically exfoliated and chemical vapor deposition (CVD)-grown MoS$_2$ flakes, yielding a high off-current ($I_{off}$)[18-19] or degenerate doping[20]. Researchers reported a high ambipolar conductivity in MoS$_2$ using ion gel-gated MoS$_2$ transistors, but the on/off ratio was only the order of 10$^2$.[21] The mechanism underlying the polar transition in MoS$_2$ needs to be explored so that simple controllable methods for achieving high-performance MoS$_2$ PFETs may be developed.[22]

**2. Experimental Details**

Multi-layer MoS$_2$ prepared using the mechanical exfoliation method was positioned on a highly doped p-Si substrate capped with 50 nm thermally oxidized SiO$_2$. A 20/40 nm Pd/Au was then deposited onto the MoS$_2$ flake to form the source and drain contacts using an electron beam evaporator. We used conventional gold chloride (AuCl$_3$) in this study as a p-type dopant [23-25] of MoS$_2$ to prepare a range of doping concentrations. The AuCl$_3$ dopant was prepared using a Schlenk line, and all operations related to the AuCl$_3$ powder or dopant were performed in a glove box to protect the reagents from the air environment. The 5 mM AuCl$_3$ dopant was spin-coated at 5500 rpm for 1 min over MoS$_2$ sample and the sample was then baked at 50°C for 5 min to get optimized non-degenerate MoS$_2$ PFET. The device performances before and immediately following doping were characterized through electrical measurements and compared. The thickness of the MoS$_2$ flakes (see Supplementary S1) used in this work was around 7–11 nm measured by atomic force microscope (AFM), and the channel length of the fabricated devices was 0.6 – 2 μm.

**3. Results and Discussion**

**Figure 1a** shows a schematic diagram of our device whereas the optical microscopy image of the device is shown in the inset of Figure 1b. The flake thickness was 10 nm. The transfer curves collected from the MoS$_2$ FET before and after doping are plotted in Figure 1b at drain biases ($V_d$) of 0.1 and −0.1 V, respectively. The pristine device displayed typical n-type behavior, as reported previously for most of the MoS$_2$ devices contacted with Pd or Au electrodes.[11, 13] After the 5mM AuCl$_3$ doping step, the transfer characteristic of the device changed to yield p-type characteristics with a low $I_{off}$ of 10$^{−13}$ A and a high $I_{on}$ on the order of μA. The on/off ratio reached 10$^7$, 100 times the value reported for MoS$_2$ PFETs, due to the high work function of MoO$_x$, as described in Ref. 17. Under a high applied $V_d$, $I_{on}$ continued to increase whereas $I_{off}$ remained constant, as shown in the Supplementary Figure S2. The output curves of the pristine NFET and doped PFET are plotted in Figure 1c. The drain current ($I_d$) was normalized by the channel width ($W$), which was 2 μm (The channel length ($L$) was 1 μm in our device.). The current density in the PFET exceeded 20 μA/μm, two orders higher than that of reported in Ref. 17. Interestingly, even if we consider the small channel length in our device, the normalized current density with respect to the channel dimensions ($I_d·L/W$) still represents the highest value yet reported among non-degenerate MoS$_2$ PFETs.[17-18, 26] Our MoS$_2$ PFET displayed Ohmic-like contact, as indicated by the linearity of the output curve shown in Figure 1c. The field effect mobilities of both the pristine NFET and the doped PFET were obtained from the transfer curves and are plotted with respect to the gate bias in Figure 1d. The highest hole mobility achieved in the doped PFETs was 68 cm$^2$/Vs, whereas the electron mobility for pristine NFET was 38 cm$^2$/Vs. Comparable hole and electron mobilities indicated that this material is suitable for use in high-performance complementary circuits.

The charge transport mechanism of our device and its transition to PFET is explained with the help of carrier injection models *i.e.* tunneling and thermionic emission. [27-29] The characteristics of Schottky emission and tunneling were discussed in detail in the Supplementary S3. **Figure 2a** shows the carrier transport path in a pristine MoS$_2$ NFET in which electrons injected from a metal electrode enter a MoS$_2$ sheet beneath the metal electrode first and are then transported to the channel. The graded color in the channel indicates varying carrier densities in different MoS$_2$ layers under gate modulation,[30] while the deep red color represents high carrier density. The Schottky barrier height associated with the Pd–MoS$_2$ contact interface has been reported to be 0.4 eV for electrons.[11] Schottky emission was found to be the main transport mechanism at subthreshold region (-5 V) proven by linear fitting in figure 2b with corresponding band diagram shown in inset. At high gate bias ($V_g$) of 15 V, Fowler-Nordheim (F-N) tunneling was expected to dominate due to thinner barrier as shown in inset of figure 2c. However, the fitting in figure 2c agreed with direct tunneling (DT). This kind of anomaly arises in devices having a small contact resistance ($R_c$) since they show linear output characteristics. If $R_c$ is a small fraction of the total resistance ($R_{tot}$), the sheet resistance ($R_s$) of the channel forms the largest contribution to $R_{tot}$ ($R_{tot} = 2R_c + R_s$),[31] and the device behaves as a simple resistor with a linear output curve, as shown in the Supplementary Figure S4b. The output curves at other $V_g$ are presented in the Supplementary Figure S4. In short, the electrons are injected from Pd electrode to MoS$_2$ channel via thermionic emission in subthreshold region and they follow F-N tunnelling at high $V_g$ mainly due to thinning of interfacial barrier. These behaviors may be understood to be the switching mechanism in a Pd-contacted n-type MoS$_2$ transistor.

Our doped MoS$_2$ PFET displayed a comparable or even better performance than the pristine NFET. The hole transport paths are described in Figure 2d, where the deep red color of top layers in the channel represents doping induced high hole concentration. The readers should note that the doping depth is in the range of 1.5 to 3 nm.[32] We inferred that holes were transported directly to the doped channel, that were extended slightly into the MoS$_2$ sheet beneath the metal contacts. Otherwise, a large potential difference will exist between the pristine n-type sheet beneath the metal contacts and the doped p-type channel, and this will induce an ultra-high $R_c$ for hole transport. It was reasonable to assume that the dopant diffused into the MoS$_2$ sheet beneath the metals since annealing was performed soon after coating the dopant. As the doped top layers had much higher hole density than the pristine n-type bottom layers, they would dominate the overall current of the PFET MoS$_2$ device. The gate dependent transfer curve can be observed due to the finite thickness of our flake (comparable to charge screening length of MoS$_2$[33]) and the non-degenerate doping concentration (will be discussed later). Bottom layers may also contribute to the hole current, however high negative bias was needed to tune the electron rich layers to hole rich one. The output curves of the doped MoS$_2$ PFET were measured and analyzed at a subthreshold gate bias (–2.5 V) and a high negative $V_g$ of –15V, as shown in Figures 2e and 2f respectively. Interestingly, we observed F-N tunneling at $V_g$ = –2.5 V and DT at $V_g$ = –15 V to be the dominant transport mechanism for hole current. It should be noted that the hole transport mechanism in the subthreshold regime is quite different from the electron transport mechanism in the same regime. Although, AuCl$_3$ doping induced hole carriers shifted the Fermi level of MoS$_2$ from the conduction band towards the valence band but the Pd Fermi level still lay on the top half of the MoS$_2$ band, inducing a triangularly shaped large Schottky barrier for holes along their interface, as shown in the inset of Figure 2e. Holes could only pass through this barrier at a high $V_d$ when F-N tunneling was significantly activated, as shown in the Supplementary Figure S4c. As the negative $V_g$ increases, the barriers thin, and $R_c$ reduces, as the result we observed DT, similar to that observed for electrons at high $V_g$ region. The corresponding band diagram is shown in the inset of Figure 2f. In brief, MoS$_2$ PFET formed due to the reduction of both $R_c$ and $R_s$ for hole transport *i.e.* reduced $R_c$ enabled the initial polar transition from n-type to p-type, while $R_s$ limited the final $I_{on}$.

The above explanation was corroborated by calculating $R_c$ and $R_s$ before and after doping in a similar thickness (8nm) device using transfer length method (TLM). An optical microscope image of the fabricated TLM patterned device is shown in the inset of Figure 3b. The channel lengths ($L$) were 0.6, 1.2, 1.8, and 2.0 μm, and the channel width ($W$) was 1.3 μm. Prior to doping, the pristine TLM device was characterized as having an $R_c$ = 7.2 $k\Omega·$μm (see the Supplementary Figure S5a). We next doped this device with 5 mM AuCl$_3$ under the conditions described above to successfully obtain a MoS$_2$ PFET, which yielded a performance comparable to previously doped 10 nm device. These results confirmed that the dopant could be controlled with good reproducibility using our method. The transfer curves measured in the doped MoS$_2$ PFET are shown in **Figure 3a**. The $R_c$ of these MoS$_2$ PFETs was calculated to be 2.3 $k\Omega·$μm, comparable to 1.8 $k\Omega·$μm obtained from a Ti contact in an n-type MoS$_2$ device,[34] while Ti is thought to be an Ohmic contact metal for MoS$_2$ NFET. The gate-dependent $R_c$ and $R_s$ values obtained from the doped MoS$_2$ PFET are plotted in Figure 3b. $R_c$ and $R_s$ dominated $R_{tot}$ at small and high negative $V_g$ respectively, consistent with our above hypothesis. (See the Supplementary Figure S5b, which describes the results obtained from the pristine NFET)

Low-temperature measurements were carried out on a 10 nm thick 2 terminal $MoS_2$ PFET device. The temperature-dependent transfer curves are shown in Figure 3c. Generally, the device current mainly depends on $R_s$ and $R_c$ i.e. $I_d=V_d/(2R_c+R_s)$, and these both resistances have different sensitivities to temperature. $R_s$ is limited by electron-phonon scattering and therefore it decreases as temperature falls. On the other hand, $R_c$ either has a very weak temperature dependence or becomes higher when the temperature is lowered due to F-N tunneling or thermionic emission respectively.[35] The increasing trend of $I_d$ with lowering temperature confirmed that $R_s$ dominated hole transport in our doped $MoS_2$ PFET under relatively high negative $V_g$. The inset of Figure 3c shows the output curve measured at 133 K, in which the linear and symmetric characteristics indicated Ohmic-like behavior. These results suggested a small $R_c$ at the metal–$MoS_2$ interface in our doped $MoS_2$ PFET. The transfer curves measured at 296 K (room temperature) and 133 K are plotted in Figure 3d, The on/off ratio increased to $10^9$ at 133 K and was two orders higher than the room temperature result of $10^7$. The increase in $I_{on}$ was attributed to reduced phonon scattering at lower temperatures, whereas the $I_{off}$ reduction was attributed to a decrease in the gate leakage. We noticed that the n-type branch appeared at lower temperatures. The origin of electron current was studied in detail in Supplementary Figure S6. Next, we calculated the hole mobilities at various temperatures as shown in Figure 3e. The mobilities were obtained at relatively high negative $V_g$, in which case $R_c$ became negligible, and thus the 2-probe measurements can be applied to get reliable mobility values. The mobility clearly increased as the temperature decreased, and a peak mobility of 132 cm$^2$/Vs was reached at 133 K. These results indicated phonon scattering, which could be fit to μ~T$^{-γ}$, where the exponent γ was equal to 0.7 for our device.[36] To the best of our knowledge, this is the first reported indication of intrinsic hole transport in an MoS$_2$ PFET, independent of $R_c$. We understand that, besides $R_c$ there can be cases that lead to incorrect mobility estimation. For example, graded carrier concentration in vertical direction induced by both gating and chemical doping may make our device system complicated,[4, 23, 37-38] and this cannot be addressed by using the conventional field effect equation. More studies are required in the future to address the mobility issue related to multi-layered 2D materials showing graded carrier concentration.

Since $R_c$ limited the polar transition of MoS$_2$ from NFET to PFET, contact engineering efforts aimed at reducing $R_c$ were implemented for achieving high-performance MoS$_2$ PFETs. Here, we applied a graphene buffer layer at Pd-MoS$_2$ interface, expecting to induce hole injection layer for MoS$_2$ since high work function of graphene is formed after AuCl$_3$ doping[39-40] and graphene work function can be further increased by effective gating.[41] A schematic diagram of the MoS$_2$ transistor prepared with a graphene buffer layer is shown in **Figure 4a**. After exfoliation of the MoS$_2$ flake (7 nm) onto a 50 nm SiO$_2$/Si substrate, two separate thin strips of graphene sheets were stacked onto the surface of the MoS$_2$ flake to form electrodes using the transfer method.[42] Next, a 20 nm Pd layer and a 40 nm Au layer were deposited onto the graphene sheets. Graphene thickness was measured to be 2 nm by AFM (Supplementary Figure S7). An optical microscope image of the fabricated device is shown in the inset of Figure 4b. Our Gr/Pd/Au contacted device resulted in normal MoS$_2$ NFET that is inconsistent with the previous study.[43] The transfer curves of Gr/Pd/Au contacted device before and after doping were shown in figure 4b. The current was normalized to the channel dimension. Normalized $I_{on}$ for the device prepared with a graphene buffer layer after doping was nearly equal to the value measured in the device prepared without a graphene buffer layer. This further strengthens our previous conclusion that $I_{on}$ is limited by $R_s$, because the doping concentrations were equal in both devices and therefore $R_s$ and $I_{on}$ were similar as well. The field-effect hole mobility was calculated to be 72 cm$^2$/Vs.

For the exploration of the superiority of graphene buffer layer in the respect of $R_c$, the effect of $R_s$ must be excluded. So we performed high concentration doping in devices with and without graphene buffer layer on the same flake (see Supplementary figure S7c for the optical microscope image). The flake thickness is 11 nm. Annealing at 100°C for 10 min was proven to result in high doping concentration for MoS$_2$ based on the doping test experiment in Supplementary S8. The red and green curves shown in Figure 4c present the normalized transfer curves of the device without graphene buffer layer doped by 5 and 20 mM dopants and annealed at 100°C for 10 min respectively. The device doped by 20 mM AuCl$_3$ provided a higher current than that of 5 mM; however, both transfer curves showed slight gate dependence. We next plotted the transfer curve (blue curve) of the counterpart device prepared with a graphene buffer layer doped with 5 mM AuCl$_3$ and annealed at 100°C for 10 min for comparison. A high current density and almost no gate dependence were observed in this device. The only difference between these two devices was the contact interface. As discussed earlier, $R_c$ dominated the total resistance of the device with highly doped channel. The graphene buffer layer provided high hole concentration at the contact interface after doping and the downward shift of its Fermi level reduced the Schottky barrier height for holes, lowering $R_c$. Thus higher performance degenerate MoS$_2$

PFET was observed in Gr/Pd/Au contacted device. The current level in these doped devices was best visualized on a linear scale, as shown in the inset of Figure 4c.

We employed graphene buffer layer and applied high concentration doping to a device (9 nm) with channel covered by 20/40 nm Pd/Au. Device schematic and its optical microscope image were shown in figure 4d and inset of figure 4e respectively. By covering the large part of channel with metal, $R_s$ remained large while $R_c$ became relatively negligible. Therefore, $R_{tot}$ for this device was mainly determined by the metal-covered part of the channel. The transfer curve of this device after doping, as shown in Figure 4e, revealed hole dominated ambipolar behavior with hole current lower than the directly doped device and a wide voltage range of off-state under gate modulation. The hole dominated conduction confirmed that reducing $R_c$ for hole can switch MoS$_2$ transistor from n-type to p-type. The low hole current however indicated that $R_s$ for hole cannot reach a sufficiently small value simply by gate modulation. The inherently high electron concentration in MoS$_2$ is also a stumbling block for achieving high performance MoS$_2$ PFET after figuring out the $R_c$ issue. Other devices prepared with a channel covering have displayed similar ambipolar behaviors with enhanced electron current due to the lack of graphene buffer layer, as shown in Supplementary Figure S9. In addition to these, we also calculated the Fermi level shift in our device after non-degenerate and degenerate doping by AuCl$_3$ dopant for reference in figure 4f. Fermi level was found to be 0.14 eV above the valence band and 0.13 eV below the valence band respectively. The calculation method was demonstrated in Supplementary S10. A comparison of MoS$_2$ PFET performances using our doping method and other methods in literatures are summarized in table 1.

Complementary device applications were explored by integrating our NFET and doped PFET through wire bonding. The device schematic and corresponding electric circuits are shown in **figure 5a** and 5b respectively (see Supplementary figure S11 for the optical microscope image of the device). Back gate was applied as $V_{in}$ and supply voltage ($V_{dd}$) was applied to the PFET. Figure 5c shows the voltage transfer characteristics of the MoS$_2$ inverter at different $V_{dd}$. Clear signal inversion is observed with high $V_{out}$ and low $V_{in}$ and *vice versa*. The dc voltage gain (g = $\partial V_{out}/\partial V_{in}$) was calculated and shown in figure 5d.

## 4. Conclusion

We demonstrated that, for MoS$_2$ as an intrinsically strong n-type semiconductor, contact engineering alone did not yield a high-performance MoS$_2$ PFET, but doping or effective channel gating was needed to improve the device performance. The introduction of a graphene buffer layer was able to reduce $R_c$ for holes. A controllable chemical doping method was used to prepare high-performance MoS$_2$ PFETs with a room temperature hole mobility of 72 cm$^2$/Vs (132 cm$^2$/Vs at 133 K), an on/off ratio exceeding 10$^7$, and a low $R_c$ of 2.3 $k\Omega\cdot\mu$m. We successfully demonstrated a MoS$_2$ CMOS inverter based on the pristine NFET and chemically doped PFET, showing the great potential of our controllable chemical doping method in the application of complementary electronic devices.

**Supporting Information**
Supporting Information is available from the Wiley Online Library or from the author.

**Acknowledgements**

This work was supported by the Basic Science Research Program through the National Research Foundation of Korea (NRF) (2013R1A2A2A01015516), and by the Global Frontier R&D Program (2013M3A6B1078873) at the Center for Hybrid Interface Materials (HIM), funded by the Ministry of Science, ICT & Future Planning.

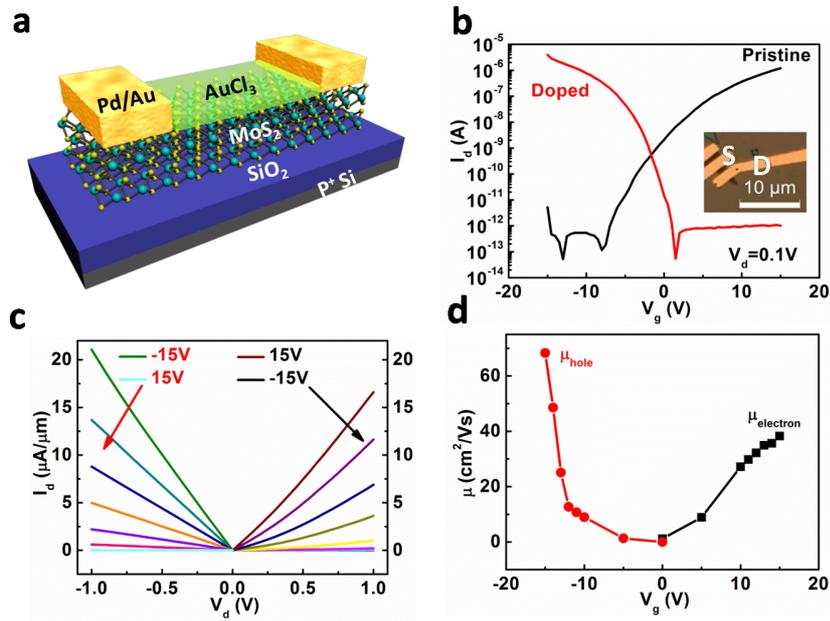

**Figure 1. Characteristics of a pristine n-type MoS$_2$ transistor and a AuCl$_3$-doped p-type device. a,** Schematic diagram showing our MoS$_2$ device and the chemical doping method. **b,** Logarithmic-scale transfer curves collected from an n-type MoS$_2$ device and its p-type transfer characteristics after AuCl$_3$ doping at $V_d$ = 0.1 V. The inset shows an optical microscope image of the device. **c**, Output curves collected from the device before and after AuCl$_3$ doping. **d**, Extracted field-effect mobility as a function of $V_g$ in a pristine n-type device or a doped p-type device.
15

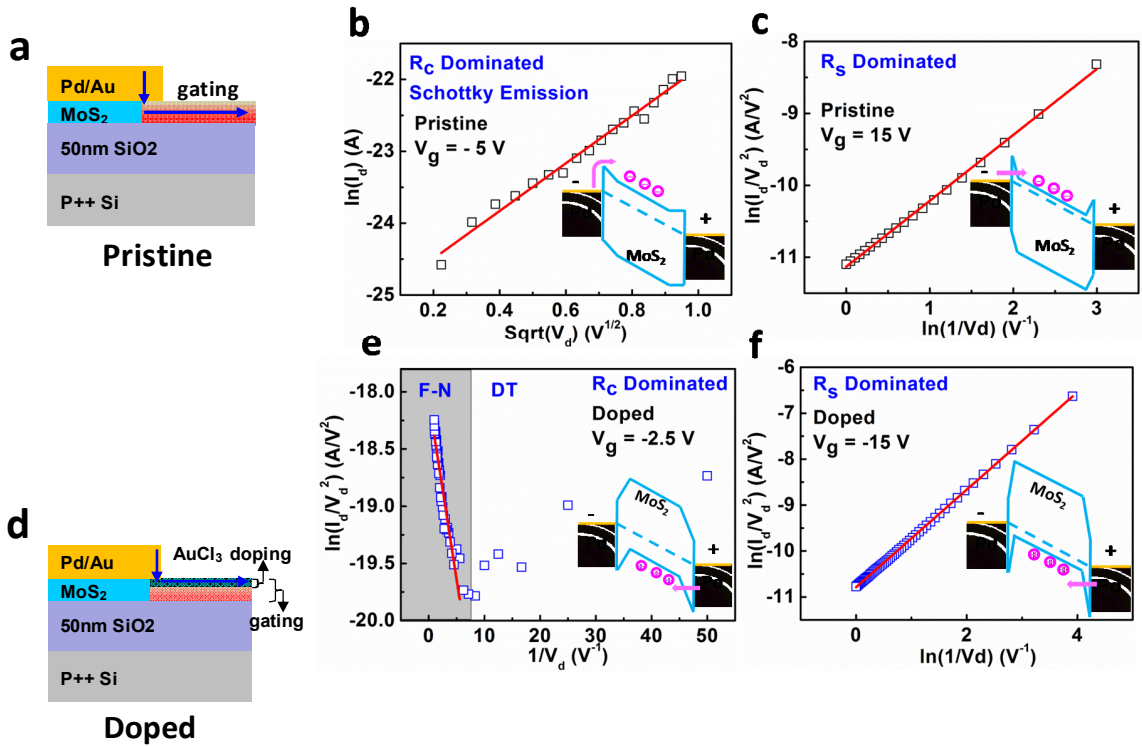

**Figure 2. Switching mechanism underlying the pristine MoS$_2$ NFET and AuCl$_3$ doped PFET operation. a** and **d,** Schematic diagrams of the carrier transport path in a pristine n-type transistor and a doped p-type transistor, the graded color indicates different carrier concentration under gate modulation and chemical doping. **b** and **c,** Schottky emission fits and direct tunneling fits to the output curves collected from a pristine n-type transistor in the subthreshold regime or in the high positive $V_g$ regime, respectively. The corresponding band diagrams are shown in the inset. **e** and **f**, The F-N tunneling fits and the direct tunneling fits for output curves collected from a doped p-type device in the subthreshold regime and in the high negative $V_g$ regime, respectively. The corresponding band diagrams are shown in the inset.

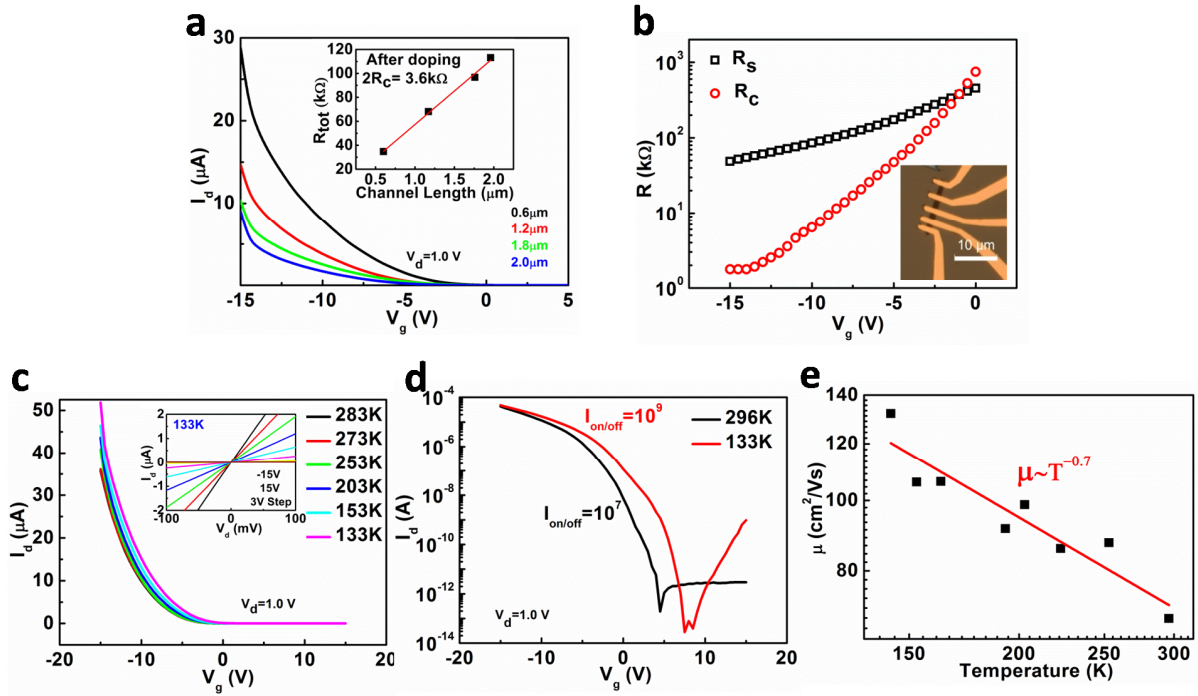

**Figure 3. TLM device and low-temperature measurements. a**, Transfer curves collected from the TLM-patterned devices prepared with various channel lengths. $V_d$ was 1.0 V. The inset shows a plot of $R_{tot}$ versus the channel length for the calculated $R_c$ value. **b**, The gate bias-dependent $R_c$ and $R_s$ values. The inset shows an optical microscope image of the TLM device. **c**, The temperature dependence of the transfer curves for a doped $MoS_2$ PFET. The output curves at 133 K are shown in the inset. **d**, Logarithmic-scale transfer curves collected from the doped PFET at room temperature or at 133 K. **e**, Temperature-dependent mobility of a PFET, and the electron–phonon scattering fit.

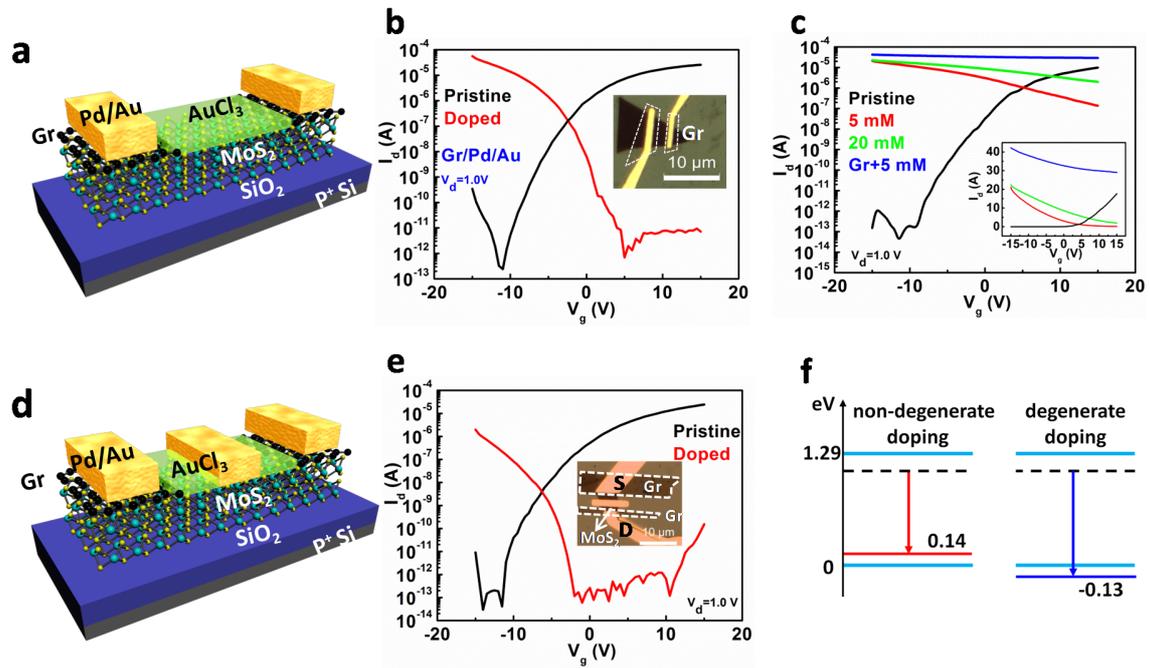

**Figure 4. Graphene buffer layer. a**, Schematic diagram showing the MoS$_2$ FET prepared with a graphene buffer layer. **b**, Transfer curve of the device before and after doping. The inset shows the optical microscope image of this device. **c**, The black line shows the transfer curve of a pristine MoS$_2$ device prepared with Pd/Au contacts. The red and green lines show the device performances after doping with 5 mM or 20 mM AuCl$_3$ dopants, respectively, after annealing at 100°C for 10 min. The blue line shows the transfer curve of a MoS$_2$ device prepared with a graphene buffer layer doped with 5 mM AuCl$_3$ dopant at 100°C for 10 min. The linear scale transfer curves are shown in the inset. **d**, Schematic diagram of an MoS$_2$ device prepared with a highly doped contact and a metal-covered channel. **e**, Transfer characteristics of the channel-covered device before and after doping. The inset shows the optical microscope image. **f**, Fermi level shift after chemical doping. All results were obtained at room temperature.

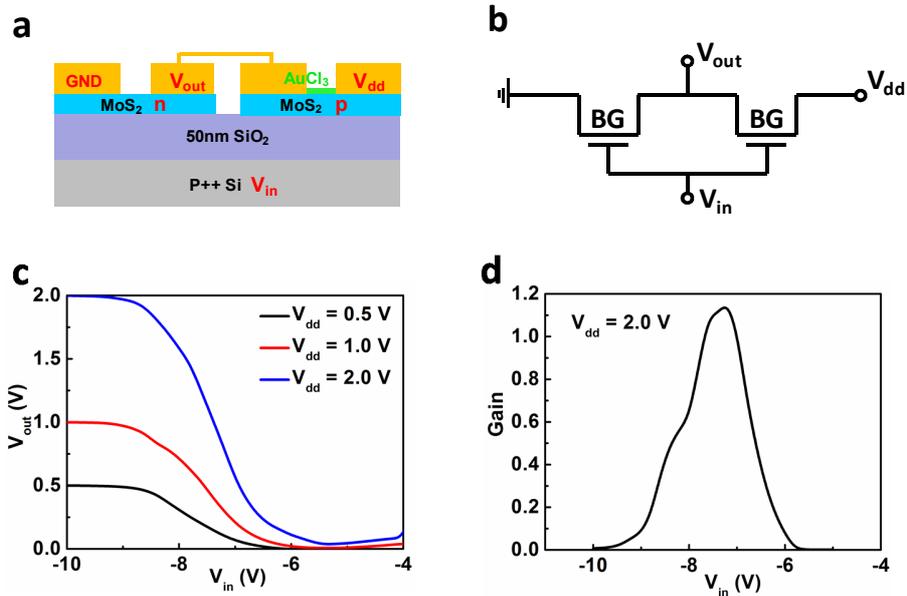

**Figure 5. MoS$_2$ inverter. a**, Schematic diagram of a MoS$_2$ inverter. MoS$_2$ PFET was prepared by chemical doping. PFET and NFET were connected by wire bonding. **b**, Electric circuit of the MoS$_2$ inverter. **c**, Voltage transfer characteristics of the MoS$_2$ CMOS inverter at different $V_{dd}$. **d**, Direct current voltage gain of the inverter at $V_{dd}$ = 2.0 V.

**Table 1**. Comparison of MoS$_2$ PFET performances enabled by different methods.

A. Non-degenerate doping

| Non-degenerate doping | Carrier density [cm$^{-2}$] | Mobility [cm$^2$/Vs] | On current density [μA/μm] | On/off ratio | Contact resistance [kΩ·μm] | Device thickness [nm] |
|---|---|---|---|---|---|---|
| **This work** | **1.5×10$^{12}$** | **68~132** | **21** | **10$^7$** | **2.3** | **7~11** |
| Plasma doping (Ref.26) | - | - | ~1 | 100 | - | 20~25 |
| MoO$_x$ buffer layer (Ref.17) | - | - | 0.13 | 10$^4$ | - | 40 |
| Nb doping (Ref.18) | ~10$^{12}$ | 6.7 | ~1 | 10$^4$ | - | 18~20 |

B. Degenerate doping

| Degenerate doping | Carrier density [cm$^{-2}$] | Device thickness [nm] |
|---|---|---|
| **This work** | **5.0×10$^{13}$** | **8** |
| Nb doping (Ref.20) | 1.8×10$^{14}$ | 61 |
| NO$_2$ doping (on WSe$_2$, Ref.8) | 2.2×10$^{12}$~2.5×10$^{12}$ | ~0.7 |

# Supporting Information

**Mechanism underlying the polar transition in chemically doped MoS₂**

*Xiaochi Liu [†], Deshun Qu [†], Jungjin Ryu, Faisal Ahmed, Zheng Yang, Daeyeong Lee, and Won Jong Yoo[*]*

X. Liu, D. Qu, J. -J. Ryu, F. Ahmed, Z. Yang, D. Lee, Prof. W. J. Yoo
Samsung-SKKU Graphene/2D Center (SSGC)
Department of Nano Science and Technology
SKKU Advanced Institute of Nano-Technology (SAINT),
School of Mechanical Engineering
Sungkyunkwan University, 2066, Seobu-ro, Jangangu, Suwon, Gyeonggi-do, 440-746, Korea.
E-mail: yoowj@skku.edu

[†]These authors contributed equally to this work

**S1 (flake thickness characterized by AFM)**

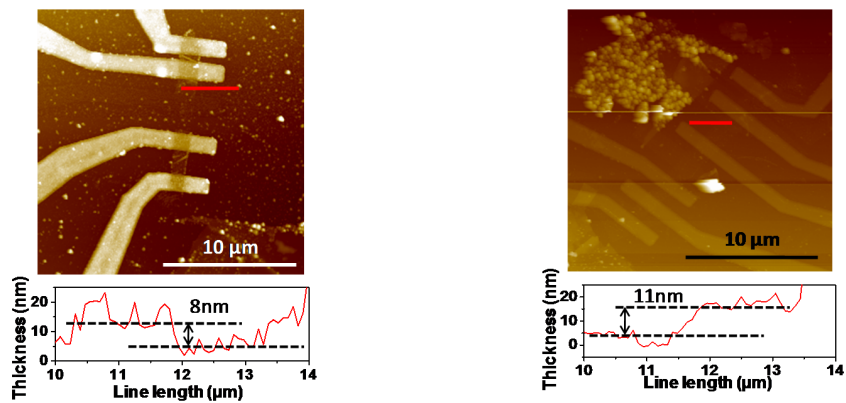

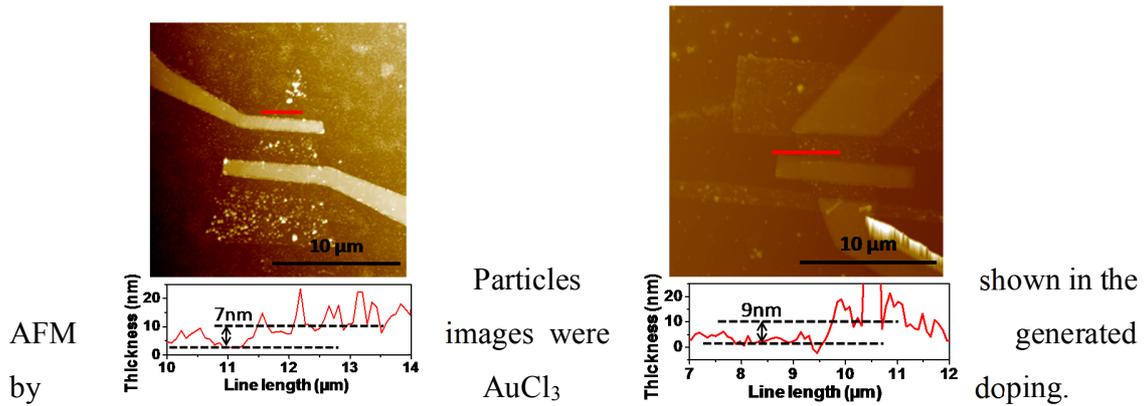

AFM images were shown in the Particles generated by AuCl$_3$ doping.

**S2 (Transfer curves of doped p-type transistor under various drain bias)**

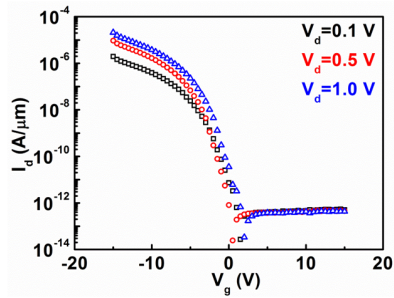

Figure S2 shows the transfer curves of doped p-type MoS$_2$ transistor at various drain bias. With drain bias increasing, the on-current ($I_{on}$) of this device keeps increasing while the off-current ($I_{off}$) stays stable at very low level of $10^{-13}$A. The channel length of our device is 1 μm, the low $I_{off}$ at 1.0 V drain bias indicates that the high current density of our doped p-type transistor does not come from the short channel effect. And the on current density can be even higher if we apply higher drain bias.

**S3 (Carrier transport mechanism at metal-semiconductor interface)**

Thermionic emission (Schottky emission), Fowler–Nordheim (F-N) tunneling, and direct tunneling are the three main mechanisms that dominate carrier transport at a metal–semiconductor interface. Thermionic emission usually dominates at a metal–MoS$_2$ interface characterized by a small Schottky barrier height, and the $I_d$–$V_d$ characteristics may be modeled using the Richardson-Schottky equation, $I_d \propto A * T^2 \exp\left[\frac{-\left(\Phi_B - \sqrt{q^3 V_d / 4\pi\varepsilon_0 \varepsilon_r d}\right)}{k_b T}\right]$, where $\Phi_B$ is the Schottky barrier height. At a certain temperature, a plot of $\ln(I_d)$ against $V_d^{1/2}$ is expected to display a linear dependence. By contrast, a high and wide barrier at a metal–semiconductor interface cannot be overcome by carriers through thermionic emission, and a high drain bias is needed to tune the barrier to a triangle shape and facilitate carrier transport via tunneling. This is the Fowler–Nordheim tunneling, the $I_d$-$V_d$ relation of which may be described according to $I_d \propto V^2 \exp\left[\frac{-4d\sqrt{2m\Phi_B^3}}{3\hbar q V_d}\right]$. A plot of $\ln(I_d/V_d^2)$ versus $1/V_d$ is expected to yield a linear dependence with a negative slope. If the barrier at the metal–semiconductor interface were small and thin, carriers could directly tunnel through it, and $I_d$ would be linearly dependent on $V_d$. In this case, a logarithm-scale plot similar to the plot used to apply the F-N tunneling model may be fit to the equation $\ln(I_d/V_d^2) \propto \ln(1/V_d)$. At a high drain bias, the carrier transport mechanism transforms into F-N tunneling from direct tunneling at a small bias.

**S4 (Output curves of pristine MoS$_2$ NFET and doped PFET under different gate bias)**

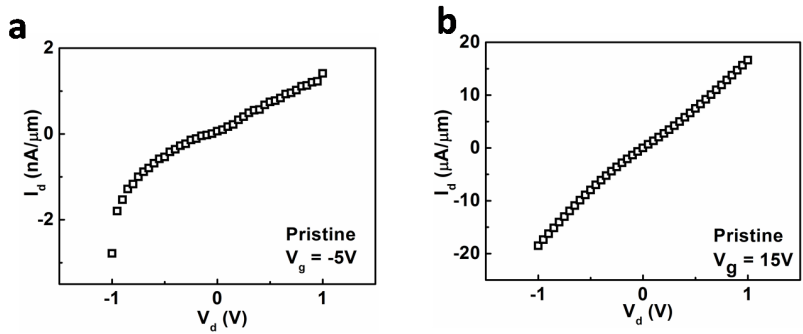

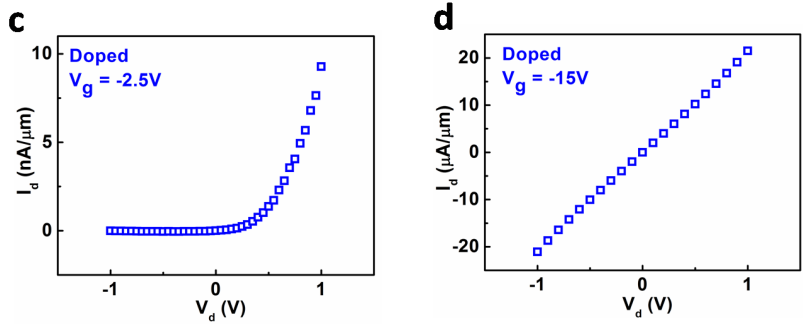

Figure S4 shows the output curves of pristine n-type device and doped p-type device at various gate biases. In the pristine undoped device, current at -5 V gate bias is just above the off-state, current starts to flow in this regime. As shown in the band diagram from the main text, there is relatively small but thick Schottky barrier for electrons at Pd-$MoS_2$ interface. Schottky emission was found to be the main mechanism for current transport. The output curve in **figure S4a** shows the corresponding rectifying behaviour from Schottky contact. Compared to this small rectifying characteristic, the output curve in figure S4c has stronger rectification effect due to the high Schottky barrier for hole conduction as we demonstrated in the main text. At 15 V and -15 V gate bias, both pristine n-type and doped p-type transistor show almost linear output curves, indicating the small contact resistance in those regimes. Since the contact resistance at 15 V in pristine n-type device is higher than that of doped p-type device at -15 V, and closer to the sheet resistance of the channel (see figure 3 in the main text), so its output curve in figure S4b is less linear than the p-type output curve in figure S4d.

## S5 (TLM measurement of pristine MoS$_2$ NFET)

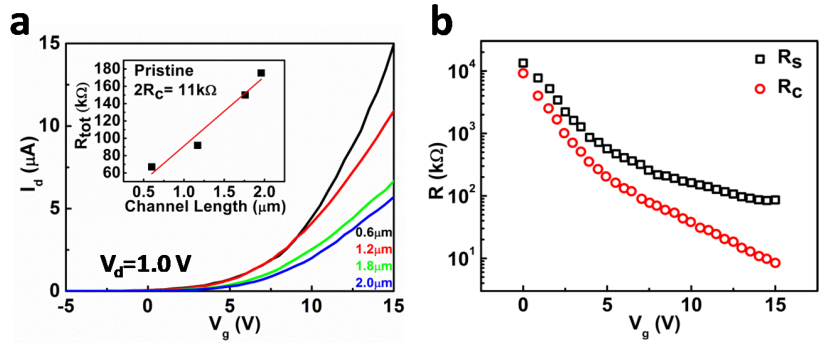

**Figure S5 a**, Transfer curves of pristine NFET TLM pattern device of various channel length. Drain bias is 1.0 V. The inset is the plot of $R_{tot}$ versus channel length for $R_c$ calculation. **b** The gate bias dependent $R_c$ and $R_s$.

**Figure S5a** is the plot of transfer curves with different channel length of pristine n-type device. $R_{tot}$ was calculated from these transfer curves at 15 V gate bias and plotted with channel length in the inset of figure 4a. A linear fit shows that 2 $R_c$=11 $k\Omega$, and the channel width was measured to be 1.3 μm, so $R_c$ is 7.2 $k\Omega·\mu m$ calculated from $R_c·W$. Gate dependent $R_c$ and $R_s$ of pristine n-type device are plotted in Figure S5b.

## S6 (Temperature dependent conduction of electrons in the doped p-type MoS$_2$ transistor)

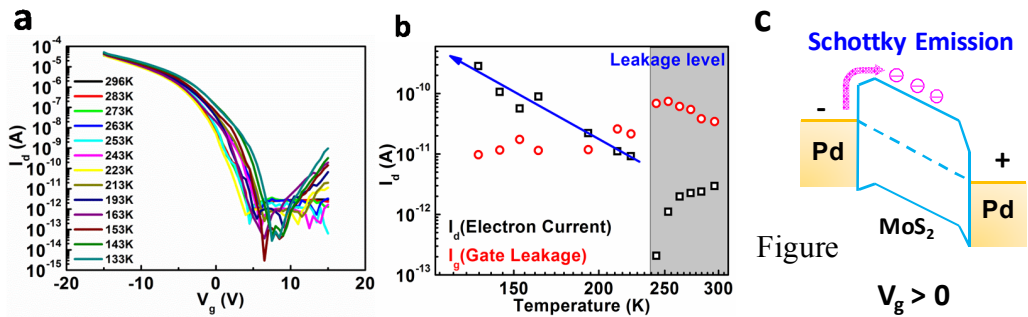

S6 **a**, Temperature dependent transfer curves of doped p-type MoS$_2$ transistor, drain bias is -1.0 V. **b**, Plot of electron current versus temperature at gate bias of 15 V. Gate leakage is shown by red circles. **c**. Band diagram for electron conduction at high positive gate bias.

The transfer curve of our doped MoS$_2$ transistor shows p-type behaviour with only hole conduction at room temperature shown by the black curve in **figure S6a**. But with temperature decreasing, electron current begins to show up and increases with lowering temperature as shown in figure S6b. Electron current is plotted by black squares, gate leakage (red circles) is also plotted for comparison. At relatively high temperature, electron current is under leakage level, corresponding to the off-state of transfer curves in figure S6a. The increasing trend of electron current with lowering temperature exclude the contact dominated carrier transport in the device. The temperature dependence may come from the sheet resistance dominated transport, which will be affected by the electron-phonon scattering. Since the channel was doped to be p-type, gate modulation is too weak to shift the Fermi level up, resulting in large sheet resistance for electron conduction. Band diagram of this regime is shown in figure S6c. Electrons are injected from the electrode by thermionic emission, although contact resistance is high due to the high Schottky barrier, total resistance is still determined by the large sheet resistance, leading to the electron-scattering dominated transport (current increases with the temperature decreasing).

**S7 (Characterization of devices with graphene buffer layer)**

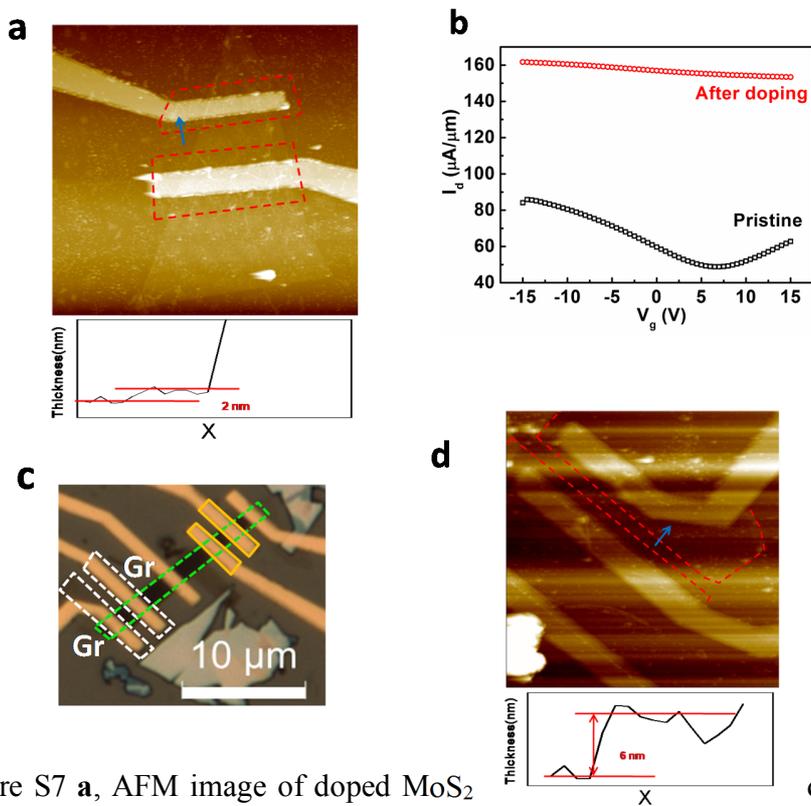

Figure S7 **a**, AFM image of doped $MoS_2$ device with graphene buffer layer (figure 4b device). **b**, Transfer curves of graphene before and after doping. **c**, Optical microscope image of devices with and without graphene buffer layer (device used in figure 4c). **d**, AFM image of graphene buffer layer contacted device from c.

**S8 (Doping process characterization)**

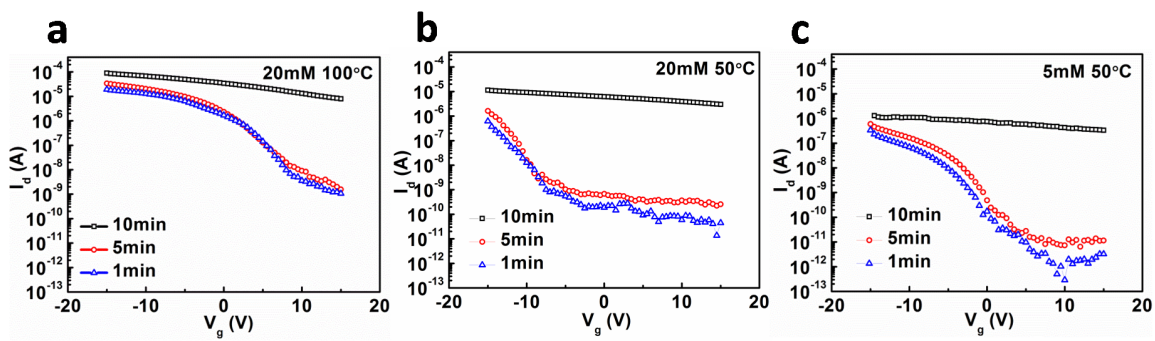

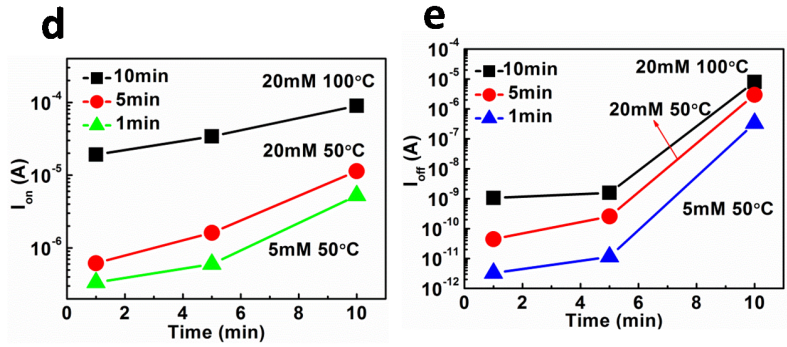

Figure S8. Doping characteristics. **a**, **b**, **c** Transfer curves of doped p-type MoS$_2$ transistor under various doping concentration and annealing temperature and time. **d** and **e** On- and off-current under different doping conditions.

Since no comparable results to our work were obtained previously despite that AuCl$_3$ is widely used p-type dopant for various materials such as graphene, MoS$_2$, and carbon nanotubes, we carry out experiment on the characterization of doping process. By varying doping concentration, annealing temperature and time, performance variation of the doped p-type device are shown in figure S8. We test 5mM, 20mM AuCl$_3$ dopant and set annealing temperature at 50 ˚C and 100 ˚C, annealing time for 1, 5 and 10 minutes. Normalized transfer curves are shown in **figures S8a**, S8b, S8c respectively. Transfer curves are normalized by channel length and width as $I_d·L/W$. They have similar trend with respect to annealing time. 10 minutes annealing results in degenerate like transfer characteristic, with current density increasing with annealing temperature and dopant concentration. 5 and 1 min annealing show similar on- and off-current level, while 5 min annealing leads to slightly higher on- and off-current than 1 min annealing. Temperature is also a very important parameter for the doped device performance. For the same dopant concentration and annealing time, high annealing temperature gives higher current density for both on-current and off-current. Current density also increases with dopant concentration, which can be concluded by comparing figure S8b and S8c. On-current and off-current are plotted with respect to dopant concentration, annealing time and temperature as shown in figure S8d and figure S8e, where clear dependence is observed. To get decent on-current density and low off-current, the doping process should be carefully controlled. 5 mM dopant, annealing at 50 °C for 5 min is a good choice as we showed in table 1 of the main text.

**S9 ($R_s$ (channel resistance) dominated device performance)**

Fig S9. **a**, Transfer 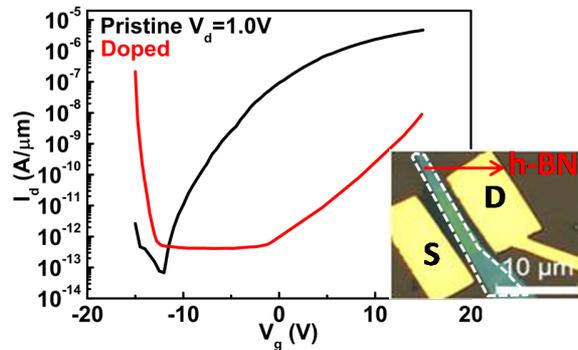 curve of pristine n-type $MoS_2$ transistor and doped ambipolar device with 2D h-BN as the mask for channel covering.

In the main text, we have shown the ambipolar behavior of device with graphene buffer layer and highly doped contact, channel is covered by e-beam deposited metal. Figure S9 shows enhanced ambipolar behavior of another device with channel masked by 2D h-BN. Since there is no graphene buffer layer at the contact, $R_c$ for holes was higher in this device, so the electron conduction branch was more clear as shown above.

**S10 (Fermi level shift calculation)**

In a 2D $MoS_2$ field effect transistor, mobility

$$\mu = \frac{1}{C} \times \frac{d\sigma}{dV_g}$$

Where $C$ is the gate capacitance, $C = \frac{\varepsilon}{t} = \frac{3.9 \times 8.85 \times 10^{-12} F/m}{50 \times 10^{-9} m} = 6.9 \times 10^{-4} \, F/m^2$, for 50 nm SiO$_2$, $V_g$ is the back gate bias, $\sigma$ is the conductivity calculated as:

$$\sigma = \frac{L}{W} \times \frac{I_d}{V_d}$$

Where $L$ and $W$ are channel length and width respectively, $I_d$ is the drain current, $V_d$ is the drain bias. The conductivity of a semiconductor is defined as:

$$\sigma = ne\mu$$

Where $n$ is the number of charge carrier per unit area, e is the elementary electrical charge. So we can get

$$n_{2D} = \frac{I_d L}{eWV_d \mu}$$

On the other hand, carrier density in a electron or hole dominated semiconductor is calculated by the following integration:

$$n_{2D}(n) = \int_{E_C}^{\infty} D(E) f(E) \, dE$$

$$n_{2D}(p) = \int_{-\infty}^{E_V} D(E) \big(1 - f(E)\big) \, dE$$

Here $D(E)$ is the 2D density of states[1] :

$$D(E) = \frac{2m^*}{\pi \hbar^2}$$

$m^*$ is effective mass of the carriers. Electron and hole effective mass are $0.45m_0$ and $0.43m_0$ respectively from Ref [2]. $m_0$ is the electron rest mass.

*f(E)* is the Fermi-Dirac distribution function :

$$f(E) = \frac{1}{e^{(E-E_f)/kT} + 1}$$

After the integration, the variable *E* will be eliminated, and we can get equations about $E_c$-$E_f$ and $E_v$-$E_f$ from the electron and hole carrier dominated devices respectively. Since $n_{2D}$ can be obtained from the electrical measurement of the device, we can finally get the Fermi level of the device under various gate biases.

**S11 (Optical microscope image of inverter)**

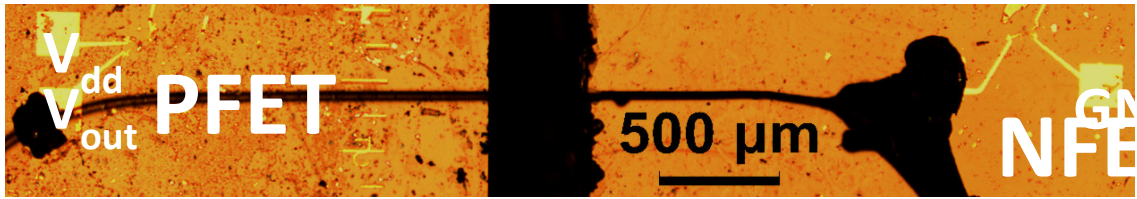